\documentclass[11pt]{article}

\let\ssection=\section
\renewcommand{\section}{\setcounter{equation}{0}\ssection}

\newcommand{\half}{{\scriptstyle{\frac{1}{2}}}}

\def\parag{\hfil\break} 
\def\kikezd{\parag\underbar}
\def\p{{\partial}}

\def\vp{{\vec p}}

\newcommand{\FS}{F\cdot S}

\begin{document}

\setlength{\baselineskip}{16pt}

\title{Anyons with anomalous gyromagnetic ratio
\& the Hall effect}

\author{ 
 C.~Duval 
 \\ 
 Centre de Physique Th\'eorique, CNRS\\ 
 Luminy, Case 907\\ 
 F-13288 MARSEILLE Cedex 9 (France)\\ 
 (UMR 6207 du CNRS associ\'ee aux 
 Universit\'es d'Aix-Marseille I et II \\ 
 et Universit\'e du Sud Toulon-Var; laboratoire 
 affili\'e \`a la FRUMAM-FR2291) 
 \\[8pt] 
 P.~A.~Horv\'athy 
 \\ 
 Laboratoire de Math\'ematiques et de Physique Th\'eorique\\ 
 Universit\'e de Tours\\ 
 Parc de Grandmont\\ 
 F-37200 TOURS (France) 
 \\ 
 } 

\date{\today}

\maketitle

\begin{abstract}
      Letting the mass depend on the spin-field coupling
      as $M^2=m^2-(eg/2c^2)F_{\alpha\beta}S^{\alpha\beta}$,
      we propose a new set of relativistic planar equations of motion for
      spinning anyons. Our model can accommodate any gyromagnetic ratio
      $g$ and provides us with a novel  version of the Bargmann-Michel-Telegdi
      equations in $2+1$ dimensions. The system becomes singular
      when the field takes a critical value, and, for $g\neq2$,
      the only allowed motions are those which satisfy the Hall law.
      For each $g\neq2,0$ a secondary Hall effect arises also for
      another critical value of the field. The non-relativistic
      limit of our equations yields new models which generalize
      our previous ``exotic'' model, associated with the
      two-fold central extension of the planar Galilei group.
\end{abstract}

\noindent\texttt{hep-th/0402191} (Revised Version)

\section{Introduction}

Most theoreticians argue that the gyromagnetic ratio of anyons
must be $g=2$ \cite{CNP,anyons,anyoneq}.
Their statement is, however, contradicted by
experimentalists, who found that in a GaAs semiconductor
$g$ can be $-0.44$ \cite{Ezawa}; in the Fractional Hall Effect,
it can be close to zero \cite{Ezawa,g0}.

In this Letter we
present a classical anyon model with arbitrary gyromagnetic
ratio~$g$. Our clue is that requiring
proportionality between momentum and velocity
is {\it not} mandatory, but a mere {\it assumption}
that can be relaxed in a perfectly consistent manner \cite{Papa, JMS}.

Our model, consistent with first principles,
is derived in two, independent ways.
Firstly, we formulate it within Souriau's version
of symplectic  mechanics,
equivalent to both the Lagrangian and Hamiltonian formalisms \cite{SSD}.
Letting the mass depend on
the coupling of spin to the electromagnetic field
provides us with a model valid for any $g$;
momentum and velocity are only parallel for $g=2$.
Our second approach follows Souriau's {\it Principe de
Covariance G\'en\'erale}, where
the equations of motion of a particle arise from the requirement
of covariance w. r. t. gauge transformations \cite{JMS}.

For the ordinary value $g=2$  previous results
\cite{CNP,anyoneq} are  recovered;
new  physics arises for $g\neq2$, though.
The most interesting physical application of anyons
concerns indeed the Hall effect \cite{Ezawa,QHE}, which is also
the main application of our model.
Firstly, for a certain critical value
of the field, (\ref{1vanish}) below, our system becomes singular and
{\it the only allowed motions} are those which {\it follow the Hall
law.} A {\it secondary Hall effect} arises  for another
critical value of the field, cf. (\ref{2vanish}).
Let us insist that, in both cases, the Hall effect becomes
{\it mandatory}.

Similar behaviour has been
observed before for an  ``exotic'' particle \cite{DH}, associated
with the two-fold central extension of the planar Galilei group
\cite{centralex} and
related to noncommutative mechanics \cite{NCQM, HPA}.
The free exotic model was rederived by Jackiw and Nair (JN)
as a subtle non-relativistic (NR) limit of the anyon \cite{JaNa}.
Their clue is to relate the second extension invariant~$\kappa$
to relativistic spin, $s$, by the ``magic Ansatz'' \cite{JaNa,DHspin}
\begin{equation}
     {s}/{c^2}=\kappa.
      \label{magic}
\end{equation}
Below, we extend these results to models interacting with an
electromagnetic field, and present a generalized  non-relativistic
``exotic'' particle with any $g$. 
For $g=0$, it reduces to our previous model
in \cite{DH}. Both types of Hall effects are retained
in the NR limit.
\goodbreak

\section{Anomalous anyons: a model with any $g$}\label{Ourmodel}

Now we present a whole family of equations valid for any  value
of the gyromagnetic ratio $g$.\footnote{
Greek indices $\alpha, \beta$, etc. range from $0$ to $2$
unless otherwise specified. Latin indices $i, j$ range
from $1$ to $2$. We use the metric diag$(c^2,-1,-1)$.}
   We first recall Souriau's group theoretical construction for the
classical model which underlies geometric quantization
which yields in turn the quantum representation
   \cite{SSD}. Let us consider the neutral component of
Poincar\'e group in $2+1$ dimensions
parametrized by the $2+1$-dimensional
Minkowski space vector $x^\alpha$, augmented by
the three Lorentz vectors
$U^\alpha, I^\alpha, J^\alpha$ such that the only nonvanishing
scalar products are $U_{\alpha}U^{\alpha}=c^2$,
$I_{\alpha}I^{\alpha}=J_{\alpha}J^{\alpha}=-1$.
The group has two Casimirs,
$m$ and $s$, and a free massive spinning particle is described
by the Cartan $1$-form \cite{SSD}
\begin{equation}
    \alpha_{0}=mU_{\alpha}dx^\alpha+sI_{\alpha}dJ^\alpha.
    \label{Cartanform}
\end{equation}
Then the classical motions are the projections onto Minkowski space
of characteristic curves of the kernel of $\sigma_{0}=d\alpha_{0}$.
Minimal coupling to an external electromagnetic
field amounts to adding ($e$ times) the electromagnetic two-form
$F=\half F_{\alpha\beta}dx^\alpha\wedge dx^\beta$ to $\sigma_{0}$,
\begin{equation}
      \sigma=d\left(mU_{\alpha}dx^\alpha+sI_{\alpha}dJ^\alpha\right)
      +\half eF_{\alpha\beta}dx^\alpha\wedge dx^\beta.
     \label{mincoupling}
\end{equation}


$\bullet$
The spin tensor $S^{\alpha\beta}=s(I^\alpha J^\beta-I^\beta
J^\alpha)$  satisfies the relation
$S_{\alpha\beta}S^{\alpha\beta}=2s^2$ and the constraint
$
S^{\alpha\beta}U_{\beta}=0.
$
Therefore $S_{\alpha\beta}=
s\epsilon_{\alpha\beta\gamma}U^{\gamma}$.
Introducing the shorthand
$
\FS=-F_{\alpha\beta}S^{\alpha\beta},
$
our clue is to {\it replace the (constant) bare mass $m$
in (\ref{mincoupling})
by a mass $M$ which depends on the electromagnetic field} \cite{JMS},
namely as
\begin{equation}
M^2=m^2+\frac{g}{2}\,\frac{e\FS}{c^2}
\label{newmass}
\end{equation}
provided $M^2\geq0$.
We emphasize that our procedure is consistent with the
general principles of Hamiltonian mechanics, as the
two-form (\ref{mincoupling}) is closed.
Our approach is therefore equivalent to having a Lagrangian
or, alternatively, a Hamiltonian framework. Let us also note that
our mass formula (\ref{newmass}) also
yields the Bargmann-Michel-Telegdi (BMT)~\cite{BMT}
equations in $3+1$ dimensions~\cite{JMS}.
See also Section~\ref{massformula}.

Introducing the momentum\footnote{We stress that
our $U_{\alpha}$ is the (normalized) {\it momentum} and
{\it not} the velocity, see below.} $p^\alpha=MU^\alpha$
and, hence, the spin tensor
$S_{\alpha\beta}=({sc}/{\sqrt{p^2}})\epsilon_{\alpha\beta\gamma}p^{\gamma}$
yields the Poisson brackets
\begin{eqnarray}
      \big\{x^{\alpha},x^{\beta}\big\}&=&
      -\displaystyle\frac{1}{(p^2)^{3/2}D}\,S^{\alpha\beta},\hfill
       \label{xxPois}
      \\[5pt]
      \big\{p_{\alpha},x^{\beta}\big\}&=&\delta_{\alpha}^{\beta}-
      \displaystyle\frac{e}{p^2D}\,
      F_{\alpha\gamma}S^{\gamma\beta},\hfill\label{xpPois}
      \\[5pt]
      \big\{p_{\alpha},p_{\beta}\big\}&=&-\displaystyle\frac{e}{D}
      F_{\alpha\beta}\hfill\label{ppPois}
\end{eqnarray}
where we put $D=1+{e\FS}/{2p^2}$.
Our system is regular provided  $D\neq0$.

$\bullet$ The two-form $\sigma$ lives indeed on the
unit-tangent bundle, $U_{\alpha}U^\alpha=c^2$,
of $2+1$-dimensional
Minkowski space. The momenta are coordinates
on this bundle, which can  be viewed hence as the $5$-dimensional surface
sitting in $6$-dimensional phase space defined by the constraint
\begin{equation}
      p^2=M^2c^2
      \label{Mconstr}
\end{equation}
with the mass $M$ given in (\ref{newmass}).
Assuming, for simplicity, that the electromagnetic field is constant,
a straightforward calculation shows that our restricted two-form reads
\begin{equation}
      \sigma=dp_{\alpha}\wedge dx^\alpha
      +\frac{(s/c^2)}{2M^3}\epsilon_{\alpha\beta\gamma}p ^\alpha
      dp^{\beta}\wedge dp^{\gamma}
      +\half eF_{\alpha\beta}dx^\alpha\wedge dx^\beta.
      \label{oursigma}
\end{equation}

Working with the Hamiltonian system given by the Poisson
brackets (\ref{xxPois}-\ref{xpPois}-\ref{ppPois}) and the Hamiltonian
$H=p^2-M^2c^2$ is equivalent to finding the kernel of
the closed two-form  $\sigma$ in (\ref{oursigma}).
Generically, i. e when
\begin{equation}
     D=1+\frac{e\FS}{2M^2c^2}
     \label{factor1}
\end{equation}
does not vanish, the kernel is $1$-dimensional, spanned by
$\delta x^\alpha$ and $\delta p^\alpha$ such that\footnote{The
complicated form of the coefficients here, and also in
(\ref{approxcoeff}) of Section \ref{NRlimit}, is due to the particular
form of the mass relation (\ref{newmass}),
see  Eq. (\ref{genvit}) in the Conclusion.}
\begin{eqnarray*}
D\,
\delta{x}^\alpha
=
\frac{p_{\beta}\delta{x}^\beta}{Mc^2}
\left[
G\,
\frac{p^\alpha}{M}
-
\frac{es}{2M^2}\left(1-\frac{g}{2}\right)
\epsilon^{\alpha\beta\gamma}F_{\beta\gamma}
\right],\label{kernel1}
\qquad
\delta p^\alpha=eF^\alpha_{\ \beta}\,\delta x^\beta\label{kernel2}
\end{eqnarray*}
where
\begin{equation}
      G=1+\frac{g}{2}\cdot\frac{e\FS}{2M^2c^2}.
      \label{factor2}
\end{equation}

Let us  first assume that neither of the factors $D$ and  $G$ vanishes.
The integral curves of  ker($\sigma$) are conveniently  parametrized
by $\tau$ such that
$
\delta \tau=p_{\alpha}\delta x^\alpha/Mc^2
$
and identified with time.
Then, we end up with
\begin{eqnarray}
D\,  \displaystyle\frac{dx^\alpha}{d\tau}
&=&
G\, \frac{p^{\alpha}}{M}
+(g-2)\frac{es}{4M^2}
\epsilon^{\alpha\beta\gamma}
F_{\beta\gamma},\label{vitesse}
\\[8pt]
\displaystyle\frac{dp^{\alpha}}{d\tau}
&=&e\,
F^\alpha_{\ \beta}\displaystyle\frac{dx^{\beta}}{d\tau}.
\label{Lorentzforce}
\end{eqnarray}

These are the new equations of motion we propose
for a relativistic particle with spin and magnetic moment
(we identify with anomalous anyons),
moving in the plane in a constant external electromagnetic field.
The Lorentz equation retains its usual form and it is only the
relation between the velocity and the momentum,
(\ref{vitesse}), which is modified. In general, the motion
depends also on the spin.
\goodbreak

Let us analyse our equations (\ref{vitesse}-\ref{Lorentzforce})
in some detail.

$\bullet$
In the absence of an external field, our construction reduces to that
of Souriau \cite{SSD}, and we recover
the free spinning anyon \cite{anyons}.

$\bullet$
Contracting (\ref{vitesse})  by the field $F_{\alpha\beta}$
and using the Lorentz equation (\ref{Lorentzforce}) yields
furthermore that
the spin-field dependent mass, $M$ in  (\ref{newmass}),
is a {\it constant of the motion},
\begin{equation}
      \displaystyle\frac{dM}{d\tau}=0.
      \label{masscons}
\end{equation}

$\bullet$
For $g=2$ the term proportional to $g-2$ in (\ref{vitesse}) drops out,
leaving us with the spin-independent equation
\begin{equation}
\displaystyle\frac{dx^\alpha}{d\tau}=
\frac{p^{\alpha}}{M\ }.
\label{g2}
\end{equation}
It follows that our parameter $\tau$ is now  proper time, since
$(dx/d\tau)^2=c^2$.
Redefining time\footnote{Time redefinition changes the
gyromagnetic factor, confirming  that
$g=2$ can be viewed as a gauge condition  \cite{Jackg},
namely that of world line
reparametrization.} according to
$
\lambda=({m}/{M})\tau
$
transforms our equations into the form posited by Chou
et al. \cite{CNP},
\begin{eqnarray}
   \frac{d{x}^{\alpha}}{d\lambda}=\frac{p^{\alpha}}{m},
\qquad
\qquad
   \frac{dp^{\alpha}}{d\lambda}=
   \frac{e}{m}F^\alpha_{\ \beta}\,p^{\beta}.\label{CNPvelmom}
\end{eqnarray}
These equations  are
associated with the two-form (\ref{oursigma}),
where $M$ is our (\ref{newmass}) with  $g=2$, and the Hamiltonian
\begin{equation}
H=
\frac{1}{2m}\big(p^2-m^2c^2\big)-\displaystyle\frac{e}{2m}\FS
=\frac{1}{2m}\big(p^2-M^2c^2\big).
\end{equation}
This latter is chosen so as
to cancel the effect of the spin term in the two-form and
to enforce the relation (\ref{CNPvelmom})
posited between $p^{\alpha}$ an $d{x}^{\alpha}/d\lambda$.

$\bullet$
The new feature of our equations  (\ref{vitesse}-\ref{Lorentzforce})
is that for  $g\neq2$  momentum and  velocity are no longer parallel.
It follows that our spin constraint is in general different from
  $S_{\alpha\beta}\,d{x}^\beta/d\tau=0$, which is also used sometimes.

$\bullet$
The general equations of motion
(\ref{vitesse}-\ref{Lorentzforce}) are highly nonlinear in the field
strength $F$.
Linearizing up to higher-order terms in the quantity
$
\displaystyle\frac{e\FS}{m^2c^2}\ll1,
$
we have $M\cong m$ where $\cong$ means ``up to
higher order terms in the field $F$''.
We
end up with the novel relativistic planar BMT-type anyon equations
\begin{equation}
\begin{array}{rll}
\displaystyle\frac{dx^\alpha}{d\tau}
&\cong&
\displaystyle\frac{p^\alpha}{m}
-\frac{es}{2m^2}\left(1-\frac{g}{2}\right)
\epsilon^{\alpha\beta\gamma}F_{\beta\gamma},
\\[12pt]
\displaystyle\frac{dp^{\alpha}}{d\tau}
&=&
\displaystyle
e\,F^\alpha_{\ \beta}\frac{dx^\beta}{d\tau}.
\end{array}
\label{planarBMT}
\end{equation}

For $g=2$ we recover the equations
(\ref{CNPvelmom}).

\section{Relativistic Hall effects}\label{Halleff}

Returning to the general case the system
becomes singular  when the factor $D$ in (\ref{factor1}) vanishes,
\begin{equation}
\frac{e\FS}{2M^2c^2}=-1.
\label{1vanish}
\end{equation}
  Then comparison with (\ref{newmass}) shows that the critical mass 
$M'$ is given by
\begin{equation}
    \big(M'\big)^2=\frac{m^2}{1+g}.
     \label{1critmass}
\end{equation}
Hence
$\big(M'\big)^2>0$ whenever $g>-1$ that we shall henceforth require.
Then the velocity is eliminated from the
l.h.s. of (\ref{vitesse}) leaving us with
\begin{equation}
\left(1-\frac{g}{2}\right)\left(p^\alpha-\frac{es}{2M'}
\epsilon^{\alpha\beta\gamma}F_{\beta\gamma}\right)=0.
\label{regcond}
\end{equation}

$\bullet$ In the ``normal'' case, $g=2$, the equation is  identically
satisfied. Then $D=G\neq0$ drops out from (\ref{vitesse}) before taking
the $D\to0$ limit, and (\ref{CNPvelmom}) holds true therefore
even when $D=G\to0$,
despite the fact that the closed two-form $\sigma$ becomes singular.

$\bullet$ In the anomalous case $g\neq2$, however, (\ref{regcond})
allows us to infer that
\begin{equation}
      p^\alpha=\frac{es}{2M'}
      \epsilon^{\alpha\beta\gamma}F_{\beta\gamma}.
      \label{Hallcond}
\end{equation}
Note that $G=1-g/2\neq0$.
Hence
$p^0=(es/M'c^2)B$ and $p^i=\epsilon^{ij}({es}/{M'c^2})E_{j}$.
Then (\ref{Hallcond}) implies that $\dot{p}^\alpha=0$
since the field is constant. Hence, by (\ref{Lorentzforce})
and $\det F=0$, we readily obtain $\dot{x}^\alpha=dx^\alpha/d\tau
\propto \epsilon^{\alpha\beta\gamma}F_{\beta\gamma}\propto
p^{\alpha}$.
The velocity $v^i=p^i/p^0$ {\it satisfies therefore the Hall law}
\begin{equation}
      v^i=\epsilon^{ij}\frac{E_{j}}{B}.
      \label{Hall}
      \end{equation}
\goodbreak

Remarkably, a {\it secondary Hall effect} can  also arise.
Let us  indeed require that the
coefficient of the momentum on the r.h.s. of (\ref{vitesse}) vanishes,
$G=0$, i.e.
\begin{equation}
\frac{e\FS}{2M^2c^2}=-\frac{2}{g}.
\label{2vanish}
\end{equation}
Then $D=1-2/g\neq0$ and the system is regular. The squared mass,
\begin{equation}
     \big(M''\big)^2=\frac{1}{3}m^2,
     \label{1/3mass}
\end{equation}
is always positive.
  The velocity will be again determined by the electromagnetic
field alone, namely according to
\begin{equation}
\dot{x}^\alpha=
\frac{3g}{4}\cdot\frac{es}{m^2}
\epsilon^{\alpha\beta\gamma}F_{\beta\gamma}.
\label{Hallbis}
\end{equation}
Then $v^i=\dot{x}^{i}/\dot{x}^{0}$
satisfies once again the Hall law (\ref{Hall})~!
Let us observe that the momentum has been decoupled,
and can be determined by solving (\ref{Lorentzforce}).

Note for further reference that both critical
conditions (\ref{1vanish})
and (\ref{2vanish}) link the fields and the spin, see Section
\ref{NRHalleff}.

\section{The origin of the mass formula
(\ref{newmass})}\label{massformula}

Our generalized model relies on the mass formula
(\ref{newmass}). Its origin can be explained from a rather
different viewpoint. Some time ago \cite{Papa} a set of
equations of motion for a general relativistic spinning
particle in a gravitational and electromagnetic field has been
proposed. These latter, called the Mathisson-Weyssenhoff-Papapetrou
equations, read
\begin{eqnarray}
\label{dP}
\dot{p}^\alpha
&=&
eF^\alpha_{\ \beta}\dot{x}^\beta-\half{}R(S)^\alpha_{\
\beta}\dot{x}^\beta
+\half{}M^{\beta\gamma}
\nabla^\alpha F_{\beta\gamma},\\[8pt]
\label{dS}
\dot{S}^{\alpha\beta}
&=&
p^\alpha\dot{x}^\beta-p^\beta\dot{x}^\alpha-M^\alpha_{\
\gamma}F^{\gamma\beta}
+M^\beta_{\ \gamma}F^{\gamma\alpha},
\end{eqnarray}
supplemented by the  conservation law $\dot{e}=0$.
Here $\nabla$ is the Levi-Civita connection of the metric and
$R(S)^\alpha_{\ \beta}=R^\alpha_{\mu\nu\beta}S^{\mu\nu}$,
where $R^\alpha_{\mu\nu\beta}$ is the Riemann curvature.
[We use the convention
$(\nabla_\mu\nabla_\nu-\nabla_\nu\nabla_\mu)v^\alpha=
R^\alpha_{\mu\nu\beta}v^\beta$.]
The quantities $p^\alpha$, $S^{\alpha\beta}$, $e$ and
$M^{\alpha\beta}$ here are interpreted as the linear momentum, the
(skew-symmetric) spin tensor, the electric charge, and the
electromagnetic dipole moment, respectively.
In this paper we only consider the flat case.

Equations (\ref{dP}-\ref{dS}) can be derived from the requirement of
gauge invariance (Souriau's ``{\it Principe de covariance
g\'en\'erale}'')
of the theory alone \cite{JMS}. They are universal
in that they hold independently of the relation between
momentum and velocity. To get a deterministic system, this latter has
to be specified by supplementary constraints \cite{JMS}.

Firstly, to guarantee the localizability of the particle, we require
$
S^{\alpha\beta}p_\beta=0.
$
Our particle should moreover carry no electric dipole
moment; this is expressed as
$
M^{\alpha\beta}=\chi\,S^{\alpha\beta}
$
where $\chi$ is some function, identified as
the scalar magnetic moment.
These conditions actually make the system deterministic.
Let us show that they also yield some unexpected result related to the
magnetization energy, (\ref{magenerg}) below.

It is straightforward to prove
that the scalar spin $s$, defined by
$s^2=\half\,S_{\alpha\beta}S^{\alpha\beta}$
is a constant
of the motion, $\dot{s}=0$. In the sequel we
promote  $s$ to  a constant of the system.
One thus finds that
$\dot{p}_\alpha\dot{S}^{\alpha\beta}p_\beta=0$; then (\ref{dS}) yields
$$
p_\alpha\dot{p}^\alpha p_\beta\dot{x}^\beta-p_\alpha
p^\alpha\dot{p}_\beta\dot{x}^\beta
- \chi \dot{p}_\alpha
S^{\alpha\beta}F_{\beta \gamma}p^\gamma=0
$$
and (\ref{dP}) enables us to write the latter equation as
$$
\half{}(p^2)\dot{}\;p_\beta \dot{x}^\beta
-
\half p^2 M^{\alpha\beta}\dot{F}_{\alpha\beta}
+\chi p_\alpha \dot{S}^{\alpha\beta}F_{\beta \gamma}p^\gamma=0.
$$
Some more work allows us to show that
$$
(p^2)\dot{}\;p_\beta\dot{x}^\beta
-
\chi p^2 S^{\alpha\beta}\dot{F}_{\alpha\beta}
+2\chi p^2 \dot{x}_\alpha F^{\alpha\beta}p_\beta=0.
$$
Since $F_{\alpha\beta}\dot{S}^{\alpha\beta}=-2\dot{x}_\alpha
F^{\alpha\beta}p_\beta$, we end up with
$ 
(p^2)\dot{}\,p_\beta\dot{x}^\beta
=
\chi{}\,p^2\,\big(F_{\alpha\beta}S^{\alpha\beta}\big)\dot{}.
$ 
The latter equation is consistent with
the mass being given by an
otherwise \textit{arbitrary} function of the spin-field coupling, viz.
\begin{equation}
   p^2=M^2c^2
   \quad\hbox{where}\quad
   M=M(\phi),
   \qquad\hbox{and}\qquad
   \phi=e\FS.
\label{magenerg}
\end{equation}

Let us insist that all these results hold true for any dimension of
space\-time.
In 3+1 dimensions, a similar procedure would yield the original
BMT equations \cite{BMT},
supplemented with a modified velocity-momentum relation
\cite{JMS}.

$\bullet$ Further justification of our key formula (\ref{newmass})
is obtained for  spin $1/2$ field with $g=2$
  by  
considering  the Dirac equation
\begin{equation}
     \Big(iD_{\alpha}\gamma^\alpha-mc\Big)\Psi=0,
\label{Dirac}
\end{equation}
where $D_{\alpha}=\p_{\alpha}-ieA_{\alpha}$
  is the gauge-covariant derivative.
Then applying the conjugate operator on the l.h.s.  yields
\begin{equation}
     \Big(D_{\alpha}D^\alpha+M^2c^2\Big)\Psi=0,
     \qquad
     M^2=m^2+\,\frac{e\FS}{c^2}
\label{Diracnewmass}
\end{equation}
which is clearly consistent with
(\ref{Mconstr}) and (\ref{newmass}) with $g=2$.
Extension to any $g$ is considered
in  the third  reference of \cite{JMS}.

\section{The non-relativistic limit}\label{NRlimit}

Let us consider, at last, the non-relativistic limit of
the general system (\ref{vitesse}-\ref{Lorentzforce}).
We shall use
$$
\FS=\frac{2s}{mc^2}\Big(-\epsilon^{ij}p_{i}E_{j}
-p^0B\Big)\approx-2sB
$$
where $\approx$
stands for ``up to higher order terms in $c^{-2}$'',
along with the generalized Jackiw-Nair-type Ansatz \cite{JaNa, DHspin}
\begin{equation}
s=\theta m^{2}c^2+s_{0}.
\label{genmagic}
\end{equation}
In the NR limit
\begin{equation}
     \begin{array}{lll}
     M^2\approx M^2_{NR}=
     m^2(1-g\theta eB),
\\[8pt]
     D\approx D_{NR}=
     \displaystyle\frac{1-(g+1)\theta eB}{1-g\theta eB},
\\[12pt]
G\approx G_{NR}=
\displaystyle\frac{1-(3g/2)\theta eB}{1-g\theta eB}
\end{array}
\label{approxcoeff}
\end{equation}
provided $g\theta eB\neq 1$.
Using $p^0\approx M$, the time component of (\ref{vitesse})
yields $\dot{x}^0\approx 1$ so that $\tau$  becomes
nonrelativistic time.
The equations of motion reduce therefore to
\begin{eqnarray}
     m\Big(1-(g+1)\theta eB\Big)\dot{x}^i&=&
     \Big(1-(3g/2)\theta eB\Big)\frac{p^i}{M_{NR}}
     -\left(1-\frac{g}{2}\right)me\theta\epsilon^{ij}E_{j}\label{NRvitesse}
     \\[8pt]
     \dot{p}^i&=&eE^{i}+eB\epsilon^{ij}\dot{x}_{j}
     \label{NRLorentzforce}
\end{eqnarray}
where the dot means now derivation w.r.t. nonrelativistic time.
For $g\theta eB\to1$ the system would blow up.
\goodbreak

$\bullet$
For $g=0$ we recover those equations  written in Ref. \cite{DH}.

$\bullet$
For $g=2$ both coefficients $D_{NR}=G_{NR}$ drop out as long as
$\theta eB\neq 1/3$ [for which it would reduces to $0=0$].
Then velocity and momentum become parallel
\begin{equation}
     m\dot{x}^i=\frac{1}{\sqrt{1-2\theta eB}}p^i
     \label{g2NR}
\end{equation}
($\theta eB\neq 1/2$).
Note that (\ref{g2NR}) is also the NR limit of (\ref{g2}).
When $\theta eB\to1/2$, (\ref{g2}) has no NR limit since  $M^2\to0$.

Let us observe that the two relativistic invariants $m$ and $s$,
interpreted as
relativistic mass and spin, respectively, give
rise, in the NR limit, to {\it two  pairs of nonrelativistic
invariants}, namely non-relativistic mass and internal energy,
and non-relativistic spin and exotic structure, respectively.
Equation (\ref{genmagic}) actually defines the
non-commutative parameter $\theta=\kappa/m^2$, and
  $s_{0}$ is interpreted as nonrelativistic spin \cite{DHspin}.

\section{Non-relativistic Hall effects}\label{NRHalleff}

Let us now consider the  critical cases in the NR limit
for $g\neq2$.

$\bullet$ The coefficient $D_{NR}$ of $\dot{x}$ on the l.h.s. of
(\ref{NRvitesse}) vanishes when
\begin{equation}
     B'=\frac{1}{1+g}\cdot\frac{1}{e\theta}
     \label{NR1critB}
\end{equation}
[which is just the NR limit of the first critical condition
  (\ref{1vanish})]. Then the Hall law, (\ref{Hall}) is satisfied.
(Alternatively, the NR limit of (\ref{Hallcond}) is
$p^i=e(\kappa/M')\epsilon^{ij}E_{j}$ and
$p^0=e(\kappa/M')B$.) Equation (\ref{NR1critB}) generalizes the result
found in \cite{DH}.
\goodbreak

The second critical case $G_{NR}=0$ [which is also the
NR limit of (\ref{2vanish})] requires
\begin{equation}
     B''=\frac{2}{3g}\cdot\frac{1}{e\theta},
     \label{NR2critB}
\end{equation}
as long as $g\neq0$. Insertion into
(\ref{NRvitesse}) yields again the Hall law (\ref{Hall}).
Alternatively, the NR limit of equation (\ref{Hallbis})
provides us with the same conclusion.

When $g=2$ no Hall effect arises, since  $\theta eB\to1/2$ is
inconsistent, and $\theta eB\to1/3$ is already regular.

\section{NR equations in the weak-field limit}\label{weakfieldlimit}

Further insight is gained by studying the weak-field limit of the
equations (\ref{NRvitesse}-\ref{NRLorentzforce}).
If both $g\theta eB\ll1$ and $\theta eB\ll1$, we can
neglect higher-order terms in the field and readily obtain
$M_{NR}\cong m(1-(g/2)\theta eB)$.
When $1-g\theta eB\neq0$,
  the weak field limit of  our
equations (\ref{NRvitesse}-\ref{NRLorentzforce}) retains the form
\begin{equation}
      \begin{array}{ll}
	 m^\star\,\dot{x}^i\cong p^i
     -\left(1-\displaystyle\frac{g}{2}\right)m\theta\epsilon^{ij}eE_{j},
	\\[12pt]
	\dot{p}^{i}\cong eE^{i}+eB\epsilon^{ij}\dot{x}_j.
	\end{array}
\label{NRgenexeqmot}
\end{equation}
where
\begin{equation}
     m^\star=m\Big(1-\theta eB\Big)
     \label{effmass}
\end{equation}
is the effective mass introduced in \cite{DH}.
These are our new, non-relativistic, ``exotic BMT'' equations
valid in a weak electromagnetic field for any $g$.
They can also be obtained taking the NR limit of the weak-field
relativistic equations (\ref{planarBMT}).

$\bullet$ For $g=2$ we find
\begin{equation}
    m^{\star}\dot{x}^{i}\cong p^{i}
\label{CNPlimit}
\end{equation}
supplemented with the Lorentz equation
$\dot{p}^{i}\cong eE^{i}+eB\epsilon^{ij}\dot{x}_{j}$,
which is in fact the NR limit of the system (\ref{CNPvelmom}).
To find this limit one has to use $p^2=M^2c^2$
instead of the naive condition $p^2=m^2c^2$, which
inconsistent with the model.
This is the only case when velocity and momentum are parallel.
\goodbreak

$\bullet$   When the gyromagnetic ratio {\it vanishes}, viz. $g=0$,
equations (\ref{NRgenexeqmot}) reduce to the
``exotic'' equations of motion discovered in \cite{DH}.
The latter are hence {\it not} the NR limit of the model in \cite{CNP},
cf. \cite{JaNa}.

$\bullet$
For a generic gyromagnetic factor, $g$, equations (\ref{NRgenexeqmot})
describe the motions of charged nonrelativistic
particles in the plane, endowed with both
anomalous magnetic moment and ``exotic'' structure, given by
the non-commutative parameter $\theta$ (alias Galilei invariant
$\kappa$).
A look at (\ref{NRgenexeqmot}) shows that
  the gyromagnetic ratio can only be detected
if $\theta\neq0$.
This is not a surprise, if we remember that by (\ref{magic})
the ``exotic'' structure
is a ``nonrelativistic shadow'' of relativistic spin.
The equations (\ref{NRgenexeqmot}) are Hamiltonian, with
\begin{equation}
      \omega=dp_{i}\wedge dx^i
     +\displaystyle\frac{eB}{2}\epsilon_{ij}dx^i\wedge dx^j
     +\displaystyle\frac{\Theta}{2}\epsilon^{ij}dp_{i}\wedge dp_{j},
      \qquad
      h=\displaystyle\frac{\vp{\,}^2}{2\widetilde{m}}+eV
      \label{NRgenexsympham}
\end{equation}
where $V$ is the electric potential, and
\begin{equation}
	\widetilde{m}=m\left(1-\frac{g}{2}\theta eB\right),
	\qquad
	\Theta=\displaystyle\frac{1-g/2}{1-(g/2)\theta eB}\,\theta
      \label{genexmtheta}
\end{equation}
provided $1-(g/2)\theta{}eB\neq0$.
For {\it any} $g\neq2$, we recover
hence our previously introduced ``exotic'' system in \cite{DH}
with redefined  parameters $\widetilde{m}$ and $\Theta$.
Interestingly, the effective mass remains unchanged,
$
{\widetilde{m}}^\star=\widetilde{m}(1-e\Theta B)
=m^\star.
$
The Poisson brackets of the coordinates associated
with the (singular) symplectic structure in (\ref{NRgenexsympham}),
\begin{equation}
       \big\{x_{i}, x_{j}\big\}=
       \frac{\widetilde{m}\ }{m^\star}
\,\Theta\,\epsilon_{ij}=
\Big(1-\frac{g}{2}\Big)\frac{m\ }{m^\star}\,\theta\,\epsilon_{ij},
\label{NRexPB}
\end{equation}
are nonvanishing except for  $g=2$, when the system becomes
commutative and reduces to the usual ``non-exotic'' particle
in an electromagnetic field.

$\bullet$ Our weak-field approximation would again
accommodate both types of Hall effects,
with some modified critical field values.
These are, however, {\it not} physical since the critical values
  are {\it not} weak, but rather fixed by
the conditions (\ref{NR1critB}) and (\ref{NR2critB}).
But for these values our weak-field derivation
given for (\ref{NRgenexeqmot})
becomes inconsistent.
\goodbreak

\section{Conclusion and outlook}\label{Discussion}

Our generalized anyon model with any gyromagnetic ratio $g$ relies on
the mass formula (\ref{newmass}), which for $g\neq2$ lifts the
conventional requirement that velocity and momentum should be
parallel. A justification comes the
Mathisson-Weyssenhoff-Papapetrou equations, also
derived from Souriau's {\it covariance g\'en\'erale} \cite{Papa, JMS}.
  It is worth mentioning that our mass formula (\ref{newmass})
is just one possibility, convenient in
  a weak electromagnetic field. Other choices have also been
considered \cite{JMS,Kunzle}. A general mass function $M(\phi)$, see (\ref{magenerg}),
would generalize (\ref{vitesse}) to
\begin{equation}
     \left(1+\frac{e\FS}{2M^2c^2}\right)
     \displaystyle\frac{dx^\alpha}{d\tau}
     =
     \left(1+\frac{e\FS}{M}\,\frac{dM}{d\phi}\right)
\, \frac{p^{\alpha}}{M}
-\frac{es}{2M^2}\Big(1-2c^2M\frac{dM}{d\phi}\Big)
\epsilon^{\alpha\beta\gamma}F_{\beta\gamma}
\label{genvit}
\end{equation}
which is (\ref{NRvitesse}), with gyromagnetic factor 
\begin{equation}
g=4c^2\,M\displaystyle\frac{dM}{d\phi}.
\end{equation} 
Again, when the system becomes
singular, cf. (\ref{1vanish}), or when
the momentum is decoupled, cf. (\ref{2vanish}),
all motions obey the Hall law, provided $g\neq2$. 

The ``Jackiw-Nair''  limit of our model provides us
with a non-relativistic model, (\ref{NRgenexeqmot}) for any $g$.
In the ordinary case $g=2$ one gets a commutative theory.
For $g\neq2$
the NR limits of (\ref{1vanish}) and  (\ref{2vanish})
yield two types of
critical values, (\ref{NR1critB}) and (\ref{NR2critB}), respectively.

What is the physical interpretation our two types of Hall effects~?
We do not have a definitive answer as yet.
A hint may come from the weak-field, NR picture, though.
Since for all $g\neq2$ the system can be brought into the
same form, namely that of \cite{DH}, it follows that
quantization of the primary critical case yields
the Laughlin description of the FQHE \cite{QHE}.
In particular, all wave functions belong to the lowest Landau level
 \cite{DH,HPA}. The first type of effect generalizes the one
in \cite{DH} to any $g$.
The second type effect is new, and is still somewhat mysterious;
it is related to a spontaneous decoupling of momentum.
\goodbreak

But is $g\neq2$ possible at all~?
The strategy of \cite{CNP}, for example, to prove that $g=2$, is  to
{\it posit} that anyons in an external electromagnetic field satisfy
the usual Lorentz equations, (\ref{CNPvelmom}).
The latter are only consistent with the $3+1$-dimensional 
BMT equations \cite{BMT} when $g=2$.
The same statement remains true for us~: consistency of
our general planar model 
with either the original \cite{BMT}, or
the suitably modified \cite{JMS}  BMT system requires $g=2$.

Other physical instances of singling out $g=2$, 
including unitarity in $3+1$ dimensional gauge theory,
string theory, as well as some extra gauge symmetry \cite{Jackg}
or supersymmetry \cite{Scripta}, are known.
Do these arguments force us to discard our equations
(\ref{vitesse}-\ref{Lorentzforce}) for $g\neq2$~?
We argue that {\it no}~:  consistency of the planar and the
spatial systems may {\it not} be mandatory --- just like it is
impossible to deduce fractional spin from a $3+1$-dimensional
model with half-integer spin~! These are the peculiar
properties of planar physics that allow for anomalous anyons.
Hence, there is no reason to discard
our  theory with an arbitrary $g$, as long as we consider $2+1$
dimensions as physical. 
Similarly, while supersymmmetry may
be a useful property, it can not be viewed as a fundamental
physical requirement.

What is the experimental situation~?
Band effects in a semiconductor renormalize the electron mass and
gyromagnetic factor. The band mass in GaAs is, for example, considerably
smaller (typically a few percent) than the electron mass.   
 Similarly, it is argued that the gyromagnetic factor in a
semiconductor is determined by the spin-orbit coupling
\cite{Ezawa,g0}.
These facts appears to be, at least, not inconsistent with our ideas
expressed here~: the small mass reminds one of our
vanishing effective mass condition, $m^\star=0$ in \cite{DH}.
The latter model has $g=0$.

Anyons have long been thought to play a
fundamental role to explain the (Fractional) Quantum Hall Effect;
to our knowledge, this is in fact the only
physical instance where anyons have experimentally been detected
\cite{Diptiman}.
We believe, therefore, that the Hall effect(s),  becoming
{\it mandatory} for some critical value(s) of our parameters,
provide us with a strong argument in
favor of the physical reality of anomalous anyons in general,
and for our theory in particular. 

At last, in $3+1$ dimensions,
similar ideas were put forward by Dixon \cite{Papa} and
developed in the seventies \cite{JMS,Kunzle}.
Previous work of Skagerstam and Stern \cite{Scripta}
espouses, in a Lagrangian framework, a viewpoint similar to ours here.
For $g=2$, our commutation relations 
(\ref{xxPois})-(\ref{xpPois})-(\ref{ppPois}) can be seen,
when taking into account our mass-shell condition (\ref{newmass}),
to agree with those, \# (26),  in~\cite{CNP}.
The difference comes precisely from our choosing (\ref{newmass}),
while the Authors of \cite{CNP} posit the simple, spin-independent
Lorentz equations (\ref{CNPvelmom}) (that imply $g=2$).

The elaboration of  the planar case and its application to
the Hall effect are, to our knowledge, new.

\kikezd{Acknowledgement}.
P.H. would like to thank Mikhail Plyushchay, Allan Stern and Peter Stichel
for discussions.
We have also benefited from correspondence with
Roman Jackiw, Pascal Lederer and Diptiman Sen.


\end{document}